\documentstyle[12pt]{article}

\def \eps {\epsilon}
\def \th {\theta}
\def \nonu {\nonumber}
\begin{document}
\begin{titlepage}
\title{ ON THE LIMITATIONS OF NEUTRINO EMISSIVITY
 FORMULA OF IWAMOTO }

\author{ Sanjay K. Ghosh, S. C. Phatak and Pradip K. Sahu\\
Institute of Physics, Bhubaneswar-751005, India}
\maketitle
\thispagestyle{empty}
\begin{abstract}
The neutrino emissivity from two and three flavour quark matter is
numerically calculated and compared with Iwamoto's formula. We find
that the calculated emissivity is smaller than Iwamoto's result by
orders of magnitude when $p_{f}(u)+p_{f}(e)-p_{f}(d(s))$ is
comparable with the temperature. We attribute it to the severe
restriction imposed by momentum conservation on the phase space
integral. We obtain an alternate formula for the neutrino emissivity
which is valid when the quarks and electrons are degenerate and
$p_{f}(u)~+~p_{f}(e)~-~p_{f}(d(s))$ is large compared to the
temperature.
\end{abstract}
\smallskip
\end{titlepage}
\eject

\newpage
\hspace {0.5in} It has been conjectured that dense stars may
consist of quark matter or quark matter core with neutron matter
outside  [1-5]. Although  theoretical
understanding of the properties of quark matter is not yet available,
various quark models have been used to calculate  the
equation of state of the quark matter and determine the properties of
quark stars  [6-9]. Unfortunately, it is found that the properties of quark
stars,
such as surface gravitational redshift z,moment of inertia I,
maximum mass M, radius R and pulsar periods P, are not
significantly different compared to those of neutron stars.
Therefore it is difficult to distinguish one from the other
observationally.

    On the other hand Iwamoto [10,11] has proposed that
neutrino
emissivity  ( $\eps$ )  could play a significant role in
distinguishing between quark and neutron stars because it
differs by orders of magnitude for the two. Particularly
$\eps$ for quark stars is larger by 6-7  orders of
magnitude than neutron stars which could lead to faster cooling
rate for quark stars, thus reducing their surface temperature. There
are, however, a number of other mechanisms [12,13] and modified URCA
processes [14] proposed to increase $\eps$ for neutron matter.

    Iwamoto [10] has derived the formula for $\eps$
using apparently reasonable approximations and this formula has
been widely used [8,15-17] to calculate $\eps$ for
two and three flavour quark matter. According to his formula
$\eps$ is proportional to baryon density  ( $n_B$ ) , strong coupling
constant  ( $\alpha_c$ )  and sixth power of temperature (T) for d
quark decay. For s quark decay T dependence of $\eps$ is
same as that for d decay. Furthermore
his results imply that electron and quark masses have negligible
effect on $\eps$ and s quark decay  ( in
case of three flavour quark matter ) plays a rather insignificant
role.

    In the present paper we want to report an exact numerical
calculation of $\eps$ and a comparison of our results
with the Iwamoto formula. Our results show that the Iwamoto
formula overestimates $\eps$ by orders of magnitude when
$p_{f}(u)+p_{f}(e)-p_{f}(d(s))$ is comparable with the temperature.
For reasonable values of $\alpha_{c}$ and baryon densities, this
quantity is much larger than the expected temperatures of neutron
stars ($\sim$ few 10ths of MeV) for two flavour quark matter, but is
comparable with temperature for three flavour quark matter.

The neutrinos are emitted from the quark matter through reactions
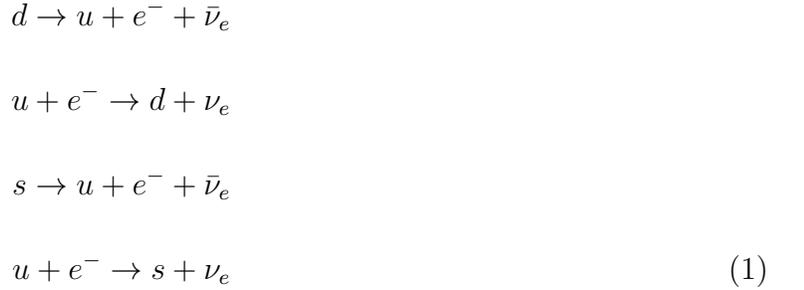
\begin{eqnarray}
d \rightarrow u + e^{-} +\bar \nu_e \nonu \\
u+ e^- \rightarrow d+ \nu_e\nonu \\
s \rightarrow u + e^{-} +\bar \nu_e \nonu \\
u+ e^- \rightarrow s+ \nu_e
\end{eqnarray}
The equilibrium constitution of the quark matter is determined by
its baryon density ($n_{B}$), charge neutrality conditions and weak
interactions given in Eq.(1). Thus, for two flavour quark matter,
\begin{eqnarray}
\mu_d = \mu_u + \mu_e   (\mu_{\nu_e}=\mu_{\bar\nu_e}=0)\nonu\\
 2 n_u - n_d -3 n_e = 0\nonu \\
n_B = ( n_u + n_d )/3
\end{eqnarray}
and for three flavour quark matter
\begin{eqnarray}
\mu_d = \mu_u + \mu_e (\mu_{\nu_e}=\mu_{\bar\nu_e}=0) \nonu \\
\mu_d = \mu_s\nonu\\
 2n_u - n_d -n_s -3n_e = 0\nonu\\
n_B=(n_u+n_d+n_s)/3.
\end{eqnarray}
The number density of species $i$ is  $n_i = g.p_f^3(i)/(6
\pi^2)$ with the degeneracy factor $g_{i}$ being two for electron and
six for quarks. For electrons
$\mu_{e}=\sqrt{p_{f}^2(e)+m_{e}^2}$ and for quarks we use [18]
\begin{equation}
\mu_q = [ {\eta \over x} + {8\alpha_c \over { 3 \pi}} ( 1- {3\over
{x \eta}} ln(x+\eta))]p_{f}
\end{equation}
where $x\equiv p_{f}(q)/m_q$ and $\eta \equiv \sqrt{1+x^2}$,
$m_q$ being the quark mass. For massless quarks Eq(4) reduces to
\begin{equation}
\mu_q=(1+ {{8 \alpha_c}\over{3 \pi}}) p_{f}(q)
\end{equation}

The neutrino emissivity $\eps$ for reactions involving
$d(s)$ quarks is calculated by using the reactions in Eq.(1). In
terms of the reaction rates of these equations, we get,
\begin{eqnarray}
\eps_{d(s)} &=&  A_{d(s)}\int d^3 p_{d(s)} d^3 p_u d^3 p_e d^3
p_{\nu} {( p_{d(s)}
. p_{\nu} ) ( p_u . p_e ) \over { E_u E_{d(s)} E_e }}
\nonu \\  &  &
\times \delta^{4}(p_{d(s)}-p_u-p_e-p_{\nu}) n(\vec p_{d(s)}) [1-n(\vec
p_u)][1-n(\vec
p_e)]
\end{eqnarray}
where $p_i=(E_i,\vec p_i)$ are the four momenta of the particles,
$n(\vec p_i)~=~ {1\over{ e^{\beta (E_i - \mu_i)} +1 }}$ are the Fermi
distribution functions and
\begin{eqnarray}
A_d = {24 G^2 \cos^2\theta_c\over{(2 \pi )^8}}\\
A_s = {24 G^2 \sin^2\theta_c\over{(2 \pi )^8}}
\end{eqnarray}
where $G$ is weak coupling constant and $\theta_c$ is Cabibbo angle.
For degenerate particles, ($\beta p_{f}(i)\gg 1$), Iwamoto has
evaluated the integrals in Eq.(5) using certain reasonable
approximations and obtained the simple expressions for $\eps_d$
 and $\eps_s$ as given below [11] .

 \begin{eqnarray}
\eps_d = {{914\over 315} {G^2 \cos^2\theta_c\alpha_c p_{f}(d) p_{f}(u)
p_{f}(e) T^6}} \nonu \\
\eps_s = {{457 \pi \over 840} {G^2 \sin^2\theta_c\alpha_c \mu_{s} p_{f}(u)
p_{f}(e) T^6}}
\end{eqnarray}
 The approximations involved in obtaining these formulas are \\
\begin{enumerate}
\item neglect of neutrino momentum in momentum conservation,
\item replacing the matrix elements by some angle averaged  value, \\
and
\item decoupling  momentum and angle integrals.
\end{enumerate}
The expressions for neutrino emissivity as obtained by Iwamoto have
been used widely. The temperature dependence of emissivity as
obtained by Iwamoto has a physical explanation. Each
degenerate fermion gives one power of T from the phase space integral
( $d^3 p_i \rightarrow {p_{f}(i)}^2 dE_i\propto T$ ). Thus one gets
$T^3$ from quarks and electrons. Phase space integral for the
neutrino gives $d^3 p_{\nu} \propto (E_{\nu}^2) dE_{\nu} \propto
T^3$. Energy conserving $\delta-$function gives one $T^{-1}$ which is
cancelled by one $E_\nu~\propto~T$ factor coming from matrix element.
So finally one gets $\eps \propto T^6$. This argument, however,
ignores the fact that
$\Delta p_{d}~(\Delta p_{s})~=~p_{f}(u)+p_{f}(e)-p_{f}(d)~(p_{f}(s))$,
which
is related to the angle between $\vec p_d$ , $\vec p_u$ and $\vec
p_e$ could be small and comparable to T. We shall demonstrate below
that precisely in this region that the Iwamoto formula fails.

Before discussing the causes of the failure of Iwamoto formula, let us
first compare our results with the Iwamoto formula
and try to find out the specific cases where the deviation is more
pronounced. In Figs.1-3 we have plotted $\eps$ vs T
for 2-flavour d decay, 3-flavour d decay and s decay respectively.
For 2-flavour d decay our results ($\eps_d$) are in good
agreement with  the emissivity calculated using Iwamoto formula
($\eps_{d(s)I}$). In Fig.1 curves (a) and (b) are $\eps_d$
and $\eps_{dI}$ respectively, for $\alpha_c$ = 0.1 and $n_B$ =
0.4. (c) and (d) corresponds to the same but for $\alpha_c$= 0.1 and
$n_b$= 1.4. It is evident from the figure that agreement of Iwamoto
results with our calculation is better for higher densities and lower
temperatures. Also $\eps_d$ is consistently smaller than
$\eps_{dI}$, the Iwamoto result, in the range of temperatures
considered. Corresponding fermi momenta of quarks and electrons are
given in Table 1. It is to be noted that all the momenta are much
larger than the temperature.

Fig.2 shows the $\eps_d$ for 3-flavour quark matter. It shows that
$\eps_{dI}$ is 2 -3 orders of magnitude higher compared to our
results.  Here ,contrary to the two flavour case, the difference
becomes more pronounced at higher densities. Fig.3 shows the
variation of $\eps_s$ with temperature. Here again it is clear
that $\eps_s$ is quite different from $\eps_{sI}$ but this
difference is less compared to that between $\eps_d$ and
$\eps_{dI}$. For all the cases the difference between our results
and those using Iwamoto formula increases at higher temperatures.
The fermi momenta of quarks and electron for 3- flavour case are
given in Table 2.  The
study of all the figures and tables above reveals that the cases
where Iwamoto formula agrees reasonably well with our results,
$\Delta p_d$ ( or $\Delta p_s$ ) is much larger than the temperature.
On the other hand when this difference is smaller or comparable with
the temperature, the Iwamoto formula overestimates the exact result
by order of magnitude. In addition to these, Fig.3 also shows that
our results are about a factor of 2.5 lower than the Iwamoto results
even at lower temperatures. We have found that this difference comes
from the approximation involved in the calculation of matrix element.

Furthermore Table 2. shows that for three flavour case electron
chemical potential ( which is same as $p_{f}(e)$ for massless
electrons ) becomes small ( $<~$1.MeV  ) for some values of $\alpha_{c}$,
$n_{B}$ and $m_{s}$. In these cases, electrons are nolonger degenerate.
Clearly, for such cases the Iwamoto formula is not applicable. This
point  is missed in earlier calculations.

Our results have profound implications on neutrino emissivity and
quark star cooling rates because all the earlier calculations have
used the Iwamoto formula and predicted large quark star cooling rates
in comparison with the neutron star cooling rates for temperatures
less than 1 MeV. Our results show that, particularly for 3-flavour
quark matter, the calculated emissivity is at least two orders of
magnitude smaller than the one given by Iwamoto formula and
therefore, the three-flavour quark star cooling rates are that much
smaller. Hence it is necessary to understand why Iwamoto formula fails.

To investigate the failure of the Iwamoto formula, we consider the integral
\begin{eqnarray}
I &=& \int \frac {d^3 p_d d^3 p_u d^3 p_e d^3p_\nu} {\eps_d\eps_u\eps_e}
\delta^{4}(p_{d(s)}-p_u-p_e-p_{\nu}) n(\vec p_{d(s)}) [1-n(\vec
p_u)][1-n(\vec p_e)].
\end{eqnarray}
Here, we have replaced the neutrino emission rate by
unity and therefore I is essentially the phase space integral.
Following the reasoning of Iwamoto, this integral should be
proportional to $T^5$. Choosing the coordinate axes such that $\vec
p_d$ is  along z-axis and $\vec p_u$ is in x-z plane and using the
3-momentum $\delta-$function to perform electron and u-quark angle
integrations, we get
\begin{eqnarray}
I = 8\pi^2\int \frac {p_d^2d p_d p_u^2dp_u p_e^2dp_e d^3p_\nu}
{\eps_d\eps_u\eps_e} [\frac
{\sqrt{1-x_u^2}}{p_ep_u(\sqrt{1-x_u^2}(p_d-p_\nu x_\nu)+
p_\nu x_\nu\sqrt{1.-x_\nu^2}\cos\phi_\nu} ] \nonu \\
\delta(\eps_{d}-\eps_u-\eps_e-\eps_\nu) n(\vec p_d) [1-n(\vec
p_u)][1-n(\vec p_e)].
\end{eqnarray}
where $x_\nu~=~\cos\th_\nu$ and $x_u~=~\cos\th_u$ is determined by
solving
\begin{eqnarray}
p_ux_u~&=&~p_d-p_\nu x_\nu~-~ [p_e^2-p_u^2(1-x_u^2)-p_\nu^2(1-x_\nu^2)
\nonu \\
{}~&~&~~~-2p_up_\nu\sqrt{(1-x_u^2)(1-x_\nu^2)}\cos\phi_\nu]^{1/2}.
\end{eqnarray}

The integral in eq(11) above is restricted to the momenta
$|p_i~-~p_f(i)|~ $ few times $T$ due to Fermi distribution
functions and the energy $\delta-$function.
Now, if we neglect the neutrino momentum in the $\delta-$functions,
we get, $x_u~=~(p_d^2+p_u^2-p_e^2)/2p_dp_u$ and the factor in the
square brackets of eq(11) becomes $1/p_dp_up_e$.

Two points should be noted at this stage.
\begin{enumerate}
\item Generally, $x_u$ is close to unity, so that $1-x_u^2$
is small. But, if $\Delta p_d$ is of the order of T,
$\sqrt{1-x_{u}^2} p_d$ can be comparable with T and $p_{\nu}$ and
therefore $p_{\nu}$ cannot be neglected in the momentum
$\delta-$functions. Particularly, the denominator in the square
bracket of eq(11) cannot be approximated by $p_ep_up_d\sqrt{1-x_u^2}$.
Thus, if $p_d\sqrt{1-x_u^2}~<~p_\nu$, one would get a power of T from
the denominator and I will not be proportional to $T^5$.
\item Secondly, the momenta may differ from the
corresponding Fermi momenta by few times T in the integral. When
$\Delta p_d~\sim~T$,  there are
regions in $p_dp_up_e-$space where $x_u~>~1$ and the rest of the
integrand is not small. Clearly, these regions must be excluded from
the integration as these values of $x_{u}$ are unphysical. If one does
not put this restriction, as is done when one factorises
angle and momentum integrals, the phase space integral will be
overestimated (and wrong) when $\Delta p_d~\sim~T$.
\end{enumerate}
The above discussion clearly shows why the integral in eq(11) should
not be proportional to $T^5$ when $\Delta p_d~\sim~T$. In order to
demonstrate this point, we have calculated the integral in eq(11)
numerically and compared with the approximation where the neutrino
momenta are neglected and the restriction imposed by $x_u$ condition
is not imposed. The calculation is done for $\alpha_c=0.1$ and for
two-flavour case. The results are shown in Fig(4). In this figure we
also show the result for a case where the electron mass is taken to
be 25 MeV. This is of course unphysical, but by adjusting the electron
mass we can reduce $\Delta p_d$. The figure clearly shows that the
approximate value of $I$ is proportional to $T^5$ where as the exact
integral is smaller than the approximate one at large T. Further
more, for 25 MeV electron mass, the departure from $T^5$ sets in at
smaller value of the temperature. This clearly shows that the
departure is dependent on the value of $\Delta p_d$. Here we would
like to mention that for some values of $\alpha_c$  and $m_{s}$
, $p_{f}(e)$ is small and is of the order of T. This
implies that electrons are no longer degenerate and deviation from
the Iwamoto result is most pronounced.

In eq(11) we have dropped the matrix element of the weak interaction
in the emissivity calculation (eq(6)). So, the discussion of
preceeding paragraphs apply to the emissivity calculation as well.
Therefore it is now clear why the Iwamoto formula
fails when $\Delta p_d$ (or $\Delta p_s$ in case of weak interactions
involving strange quarks) is close to the temperature of the quark
matter. We would like
to note that the failure of $T^6$ dependence of the emissivity
essentially arises from invalid approximations in the phase space
integration. Since similar arguments are used to obtain the neutrino
emissivity of neutron matter, it is possible that the emissivity
calculated for neutron matter may also be overestimated. We are
investigating this point.

Since the departure from the Iwamoto formula arises from the fact
that $T / \Delta p_{d}$ (or $T / \Delta p_{s}$ ) is not small, it may be
possible to fit the numerically calculated $\eps$ with a
function of the form $\eps_{I} f(x)$, where $x \ = \ T / \Delta p_{d}$
($T / \Delta p_{s}$ for strange sector ). The function $f(x)$ should be
such that for small values of $x$ it should approach unity. Choosing
$f(x) \ = \ {1.\over{1. \ + \ a x \ + \ bx^2 \ + \ cx^3}}$, we have
fitted the
calculated $\eps$ for a number of values of $n_{B}$, $\alpha_{c}$
and $ m_{s}$ and obtained the values of a, b and c. The quality of fit
is shown in Fig.5. The values of $a$, $b$ and $c$ are $-2.0$, $110.$
and $30.$ respectively. Here we would like to point out that since,
for $s$- decay, as mentioned above, there is a
difference of a factor of 2.5 in Iwamoto and our results even at
lower temperatures, the data
points for $s$- decay, in Fig.5., have been scaled accordingly.

To summarise, we have demonstrated that the Iwamoto formula of
neutrino emissivity fails when $T / \Delta p_d$ (or $T / \Delta p_s$) is
not small. The formula fails because the neglect of neutrino momentum
and factorisation of angle and momentum integrals is not valid. We
propose an alternate formula which is obtained by fitting the
numerically calculated $\eps$.

\vfill
\newpage
\centerline {\bf FIGURE CAPTIONS}
\noindent {\bf FIG.1}. Two flavour $d$-decay for $\alpha_c=0.1$;
(a) Iwamoto  results for $n_B=0.4 fm^{-3}$, (b) Our
results for $n_B=0.4 fm^{-3}$ ($\Delta p_d=6.78$),(c) Iwamoto
results for $n_B=1.4 fm^{-3}$, (b) Our results for $n_B=1.4
fm^{-3}$ ($\Delta p_d=10.29$).

\noindent {\bf FIG.2}. Three flavour $d$-decay for
$\alpha_c=0.1$ and $s$ quark mass is 150 MeV; (a) Iwamoto
results for $n_B=1.4 fm^{-3}$, (b) Our results for $n_B=1.4
fm^{-3}$ ($\Delta p_d=0.067$),(c) Iwamoto
results for $n_B=0.4 fm^{-3}$, (b) Our results for $n_B=0.4
fm^{-3}$ ($\Delta p_d=0.39$).

\noindent {\bf FIG.3}. Three flavour $s$-decay for
$\alpha_c=0.1$ and $s$ quark mass is 150 MeV; (a) Iwamoto
results for $n_B=1.4 fm^{-3}$, (b) Our results for $n_B=1.4
fm^{-3}$ ($\Delta p_s=1.613$),(c) Iwamoto
results for $n_B=0.4 fm^{-3}$, (b) Our results for $n_B=0.4
fm^{-3}$ ($\Delta p_d=9.719$).

\noindent {\bf FIG.4}. Two flavour phase space integrals for
$\alpha_c=0.1$ (a) Without restriction on $\cos\theta_u$ for
both electron mass $m_e$=0.0 and 25 MeV, (b) Exact integral for
$m_e$=0.0, (c) Exact integral for $m_e$=25 MeV.

\noindent {\bf FIG.5}. $\epsilon_{d(s)I} \over {\epsilon_{d(s)}}$
is plotted aganist $x$ where $x \ = \ {T\over {\Delta p_{d(s)}}}$.
The fitted function is $f(x) \ = \ 1.~+~ax~+~bx^2~+~cx^3$
where $a=-2.0$, $b=110.$ and $c=30.$
\vfill
\newpage

\noindent Table 1. Baryon number density $n_B$, Fermi momenta of
u-quark $p_{f}(u)$, d-quark $p_{f}(d)$, and
electron $p_{f}(e)$ for different $\alpha_c$
, where $\Delta p_d=p_{f}(u)+p_{f}(e)-p_{f}(d)$.
\vspace {0.2in}
\begin{center}
\begin{tabular}{|c|c|c|c|c|c|}
\hline
\multicolumn{1}{|c|}{$\alpha_c$} &
\multicolumn{1}{|c|}{$n_B$} &
\multicolumn{1}{|c|}{$p_{f}(u)$} &
\multicolumn{1}{|c|}{$p_{f}(d)$} &
\multicolumn{1}{|c|}{$p_{f}(e)$}&
\multicolumn{1}{|c|}{$\Delta p_d$}\\
\multicolumn{1}{|c|}{} &
\multicolumn{1}{|c|}{($fm^{-3}$)} &
\multicolumn{1}{|c|}{(MeV)} &
\multicolumn{1}{|c|}{(MeV)} &
\multicolumn{1}{|c|}{(MeV)} &
\multicolumn{1}{|c|}{(MeV)} \\
\hline
        &0.60&357.80&449.20&99.15&7.75\\
     0.1&1.00&424.22&532.52&117.56&9.20\\
        &1.40&474.57&595.79&131.51&10.29\\
\hline
        &0.60&357.71&449.25&95.43&3.89\\
     .05&1.00&424.11&532.65&113.15&4.61\\
        &1.40&474.45&595.87&126.57&5.15\\
\hline
\end{tabular}
\end{center}
\newpage

\noindent Table 2. Baryon number density $n_B$, Fermi momenta of
u-quark $p_{f}(u)$, d-quark $p_{f}(d)$, s-quark $p_{f}(s)$ and
electron $p_{f}(e)$ for different $m_s$ and different $\alpha_c$
, where $\Delta p_d=p_{f}(u)+p_{f}(e)-p_{f}(d)$ and $\Delta
p_s=p_{f}(u)+p_{f}(e)-p_{f}(s)$
\vspace {0.2in}
\begin{center}
\begin{tabular}{|c|c|c|c|c|c|c|c|c|}
\hline
\multicolumn{1}{|c|}{$m_s$} &
\multicolumn{1}{|c|}{$\alpha_c$} &
\multicolumn{1}{|c|}{$n_B$} &
\multicolumn{1}{|c|}{$p_{f}(u)$} &
\multicolumn{1}{|c|}{$p_{f}(d)$} &
\multicolumn{1}{|c|}{$p_{f}(s)$} &
\multicolumn{1}{|c|}{$p_{f}(e)$}&
\multicolumn{1}{|c|}{$\Delta p_d$}&
\multicolumn{1}{|c|}{$\Delta p_s$} \\
\multicolumn{1}{|c|}{(MeV)} &
\multicolumn{1}{|c|}{} &
\multicolumn{1}{|c|}{($fm^{-3}$)} &
\multicolumn{1}{|c|}{(MeV)} &
\multicolumn{1}{|c|}{(MeV)} &
\multicolumn{1}{|c|}{(MeV)} &
\multicolumn{1}{|c|}{(MeV)}&
\multicolumn{1}{|c|}{(MeV)}&
\multicolumn{1}{|c|}{(MeV)} \\
\hline
     &   &0.60&356.99&360.06&353.86&3.33&0.26&6.46\\
     &0.1&1.00&423.26&424.81&421.69&1.69&0.14&3.26\\
150.0&   &1.40&473.49&474.27&472.72&0.84&0.06&1.61\\
\cline{2-9}
     &    &0.60&356.99&365.89&347.62&9.28&0.38&18.65\\
     &0.05&1.00&423.26&430.29&415.98&7.33&0.30&14.61\\
     &    &1.40&473.49&479.49&467.34&6.25&0.25&12.40\\
\hline
     &   &0.60&356.99&365.93&347.58&9.70&0.76&19.11\\
200.0&0.1&1.00&423.26&429.10&417.25&6.34&0.50&12.35\\
     &   &1.40&473.49&477.66&469.25&4.52&0.35&8.76\\
\cline{2-9}
     &    &0.60&356.99&374.26&337.85&18.00&0.73&37.14\\
     &0.05&1.00&423.26&437.14&408.39&14.47&0.59&29.34\\
     &   &1.40&473.49&485.45&460.90&12.46&0.50&25.05\\
\hline
\end{tabular}
\end{center}
\end{document}